\documentclass[preprint, a4paper, showpacs, superscriptaddress]{revtex4}
\pdfoutput=1
\usepackage{graphicx}
\usepackage{amsmath, amssymb, amsfonts}

\begin{document}

\title{Anomalous Structure and Scaling of Ring Polymer Brushes}
\author{Daniel Reith}
\email{reithd@uni-mainz.de}
\affiliation{ Institut f\"{u}r Physik, Johannes Gutenberg-Universit\"{a}t\\
 D-55099 Mainz, Staudinger Weg 7, Germany}
\author{Andrey Milchev}
\affiliation{Institute of Physical Chemistry, Bulgarian Academy of Sciences, Sofia 1113, Bulgaria}
\author{Peter Virnau}
\affiliation{ Institut f\"{u}r Physik, Johannes Gutenberg-Universit\"{a}t\\
 D-55099 Mainz, Staudinger Weg 7, Germany}
\author{Kurt Binder}
\affiliation{ Institut f\"{u}r Physik, Johannes Gutenberg-Universit\"{a}t\\
 D-55099 Mainz, Staudinger Weg 7, Germany}

\begin{abstract}
A comparative simulation study of polymer brushes formed by grafting at a planar
surface either flexible linear polymers (chain length $N_L$) or (non-catenated)
ring polymers (chain length $N_R=2 N_L$) is presented. Two distinct off-lattice models are studied,
one by Monte Carlo methods, the other by Molecular Dynamics, using a fast implementation on graphics
processing units (GPUs).
It is
shown that the monomer density profiles $\rho(z)$ in the $z$-direction
perpendicular to the surface for rings and linear chains are practically
identical, $\rho_R(2 N_L, z)=\rho_L(N_L, z)$. The same applies to the pressure, exerted on a piston
at hight z, as well. While the gyration radii
components of rings and chains in $z$-direction coincide, too, and increase linearly
with $N_L$, the transverse components differ, even with respect to their scaling
properties: $R_{gxy}^{(L)} \propto N_L^{1/2}$, $R_{gxy}^{(R)} \propto
N_L^{0.4}$. These properties are interpreted in terms of the statistical
properties known for ring polymers in dense melts. 
\end{abstract}

\maketitle

\section{Introduction and motivation}

Dense layers of macromolecules grafted by special chemical groups to
non-adsorbing substrates have found enormous interest recently (for reviews see
\cite{Halperin1992,Klein1996,Szleifer1996,Grest1999,Advincula2004,Binder2011})
in view of numerous applications: colloid stabilization \cite{Napper1983};
improvement of lubrication properties \cite{Klein1996}, also in a biological
context \cite{Klein2009}, creation of functional surfaces with switchable
properties \cite{Merlitz2009}, improvement of drug biocompatibility
\cite{Storm1995}, microfluidic devices for biomolecule separation
\cite{Wang2007}, etc. The theoretical understanding of the (often unexpected)
properties of polymer brushes, due to an interesting interplay of
monomer-monomer and monomer-solvent interactions and the entropic forces resulting from the
confinement of macromolecular conformations, has been a longstanding challenge
as well (e.g.
\cite{Halperin1992,Grest1999,Binder2011,Alexander1977,deGennes1980,Milner1988,
Zhulina1991,Netz1998,Kreer2004,Galuschko2010}).

To the best of our knowledge, all this research has focused on the grafting of
linear polymers by a special end-group at one chain end; only as one exception 
the formation of ``loop brushes'' (where both chain ends are grafted to the
substrate) has been considered \cite{Yin2007}. In the latter case, although the
chains are permanently entangled (forming a ``catenated'' network-like
structure), the brush properties are found to differ at best marginally from
those of brushes with free ends \cite{Yin2007}.

In recent years, there has been a great interest in the properties of polymer
melts formed from non-catenated ring polymers
\cite{Roovers1988,McKenna1989,Gagliardi2005,Kawaguchi2006,Kapnistos2008,Nam2009,
Cates1986,Muller1996,Brown1998,
Brown1998a,Muller2000a,Muller2000,Brown2001,Hur2006,Suzuki2008,Suzuki2009,
Vettorel2009a,Vettorel2009,Hur2011,Halverson2010, Halverson2010a}. Both
synthetic polymers (such as polyethylene or polystyrene \cite{Gagliardi2005})
and biopolymers such as DNA can be prepared as closed rings, and are of great
interest as model system, to understand such diverse problems as polymer melt
dynamics when the standard reptation mechanism (that needs chain ends
\cite{Doi1986}) is eliminated \cite{Roovers1988,McKenna1989,Kawaguchi2006}, and
the organization of DNA in the cell nucleus
\cite{Vettorel2009,Kimura2007,Meaburn2007,Dorier2009, Lieberman-Aiden2009}. Note that most of the
short genomes as well as plasmids are circular \cite{Alberts2010,Witz2008} and
also actin can self-assemble into rings \cite{Sanchez2010}.
Ring polymers under various kinds of confinement are also under discussion in this
biological context (see, e.g., \cite{Liu2008,Fritsche2011}).

It is a challenging problem to understand the conformation of ring polymers
(both with respect to single collapsed rings \cite{Grosberg1988} and confined
rings \cite{Liu2008,Ostermeir2010,Fritsche2011} and rings in melts
\cite{Roovers1988,McKenna1989,Gagliardi2005,
Kawaguchi2006,Kapnistos2008,Nam2009,Cates1986,Muller1996,Brown1998,
Brown1998a,Muller2000,Muller2000a,Brown2001,Hur2006,Suzuki2008,Suzuki2009,
Vettorel2009, Vettorel2009a,Hur2011,Halverson2010,Halverson2010a}). One famous
picture is that collapsed rings (as well as rings in a melt) are ``crumpled
globules'', where each subchain of the loop is condensed in itself, and the
fractal dimension of the object is $d_f=3$. Hence the gyration radius scales
like a compact globule, $R_g \propto N^{1/3} _R$, $N_R$ being the number of
(effective) monomers in the ring polymer (hereafter, referred to as ring
length).  However, when one hypothesizes that the structure of a ring polymer in
a melt resembles a ``lattice animal''(where every bond of the lattice is
traversed by the polymer twice, in opposite directions) one predicts
\cite{Cates1986} $R_g \propto N_R^{2/5}$. The evidence from both experiment and
simulations on this issue has been discussed controversially over decades, and
only recently a resolution seems to emerge
\cite{Vettorel2009,Vettorel2009a,Halverson2010, Halverson2010a}: the ring
polymers in melts behave like lattice animals for intermediate ring lengths but
then crossover to crumpled globule-like structures for very long rings. Recall
that both powers are substantially different from the law seen for melts of
linear polymers, where a scaling like for gaussian chains occurs, $R_g\propto
N_L^{1/2}$ \cite{DeGennes1979}. We also stress that a (hypothetical) melt of ring
polymers where the rings were free to cross each other, topological
non-crossability constraints being ``switched off'', would have trivial Gaussian
statistics \cite{DeGennes1979} as well.

In the present paper, we ask what happens when ring polymers 
are grafted to a planar substrate by special 
groups at a single monomer. It is well known
that for a polymer brush formed from linear chains at large enough grafting
densities $\sigma_g$ strongly stretched polymer configurations result while in
the lateral direction the chains behave ideal \cite{Halperin1992,Binder2011}
\begin{equation} \label{eq1}
R^{(L)}_{gz} \propto \sigma^{1/3}_g N_L, R^{(L)}
_{gxy} \propto \sigma^{-1/12}_g N_L^{1/2} \quad,
\end{equation}
where $N_L$ is the number of (effective) monomers in the linear polymer. The
prefactor describing the $\sigma_g$-dependence can already be understood from
the simple Alexander-de Gennes \cite{Alexander1977,deGennes1980} blob picture,
describing a chain by a string of blobs of diameter $d \propto \sigma_g^{-1/2}$, each blob
containing $g=d^{5/3}$ monomers, taking $n=N_L/g$ steps in the
$z$-direction perpendicular to the substrate while the typical excursion parallel to the substrate
is of order $\sqrt{n} d$. The self-avoiding walk statistics inside a blob, $d\propto g^\nu$ with
$\nu=3/5$ is, of course, implicitly included in this argument. By comparing our simulation results for ring polymers with ring lengths
$N_R = 2\ N_L$ of otherwise identical linear polymers
forming a brush, at $\sigma_{g,R} = \sigma_{g,L} /2$ (to keep the number of
effective monomers strictly the same), we are able to clarify the effect of the
topological interaction on the rings in this situation. We shall show indeed
certain characteristic differences between ring brushes and linear brushes,
which can be traced back to these topological effects.

\section{Models and simulation techniques}
\begin{figure}
\includegraphics{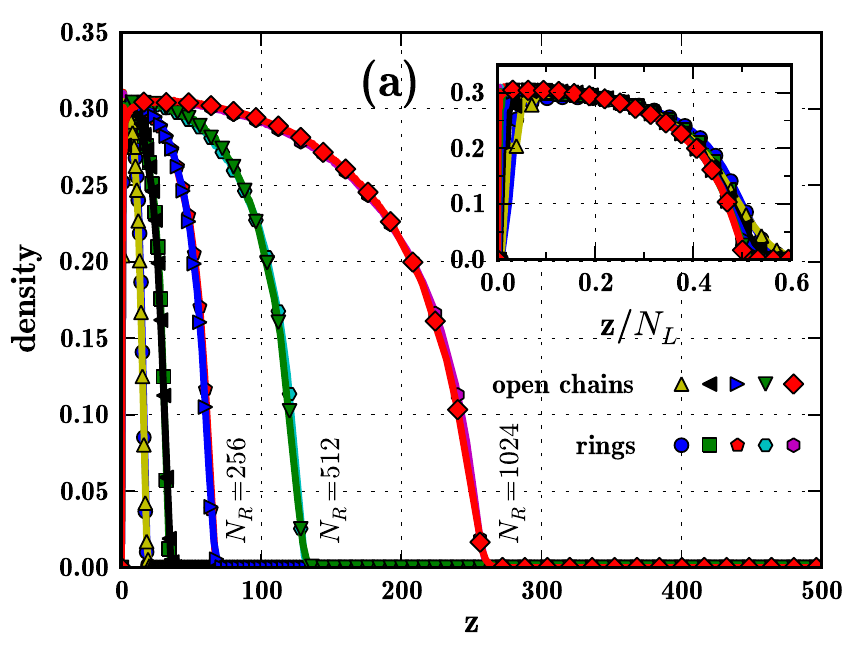}
\includegraphics[width=8cm]{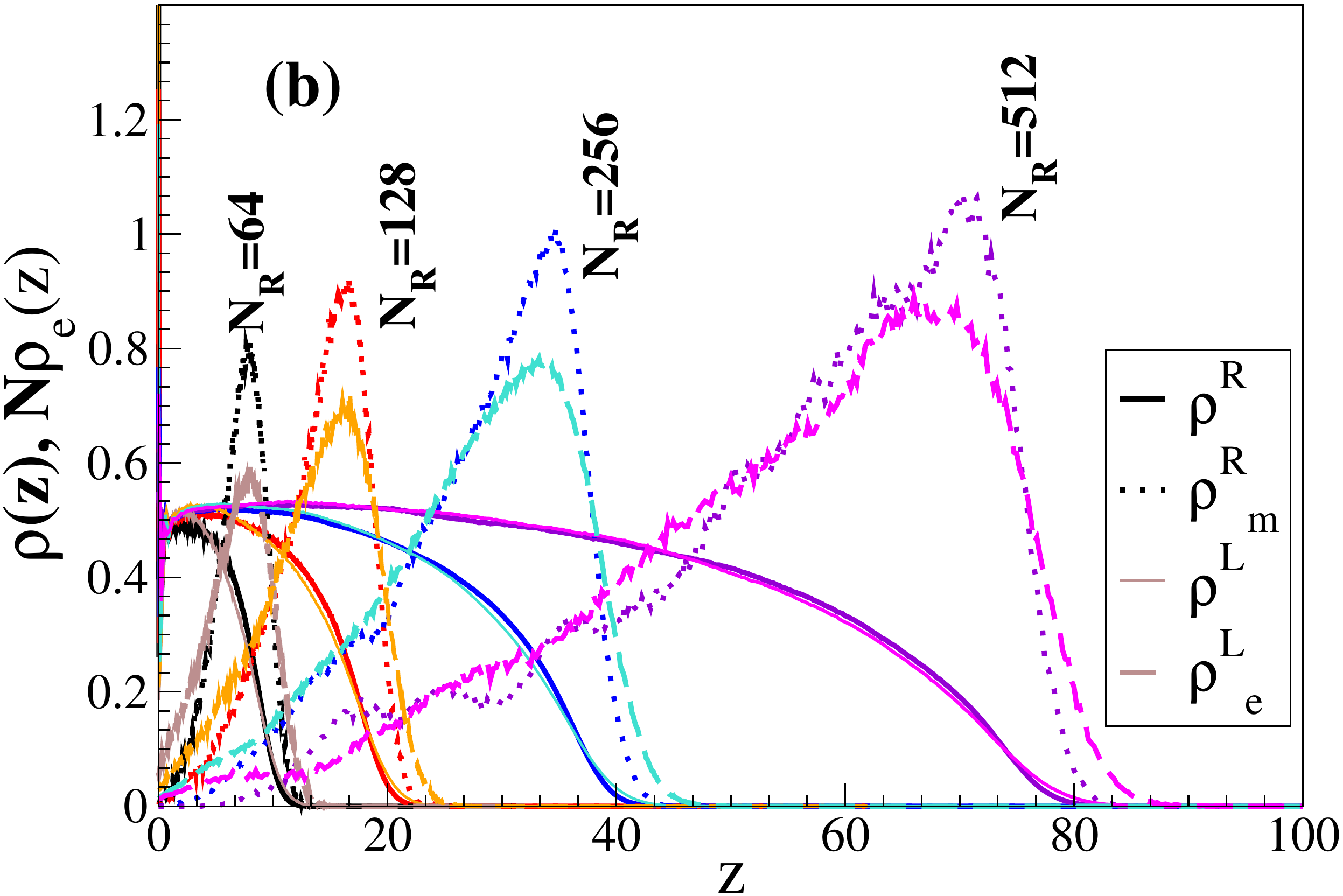}
\includegraphics[width=8cm]{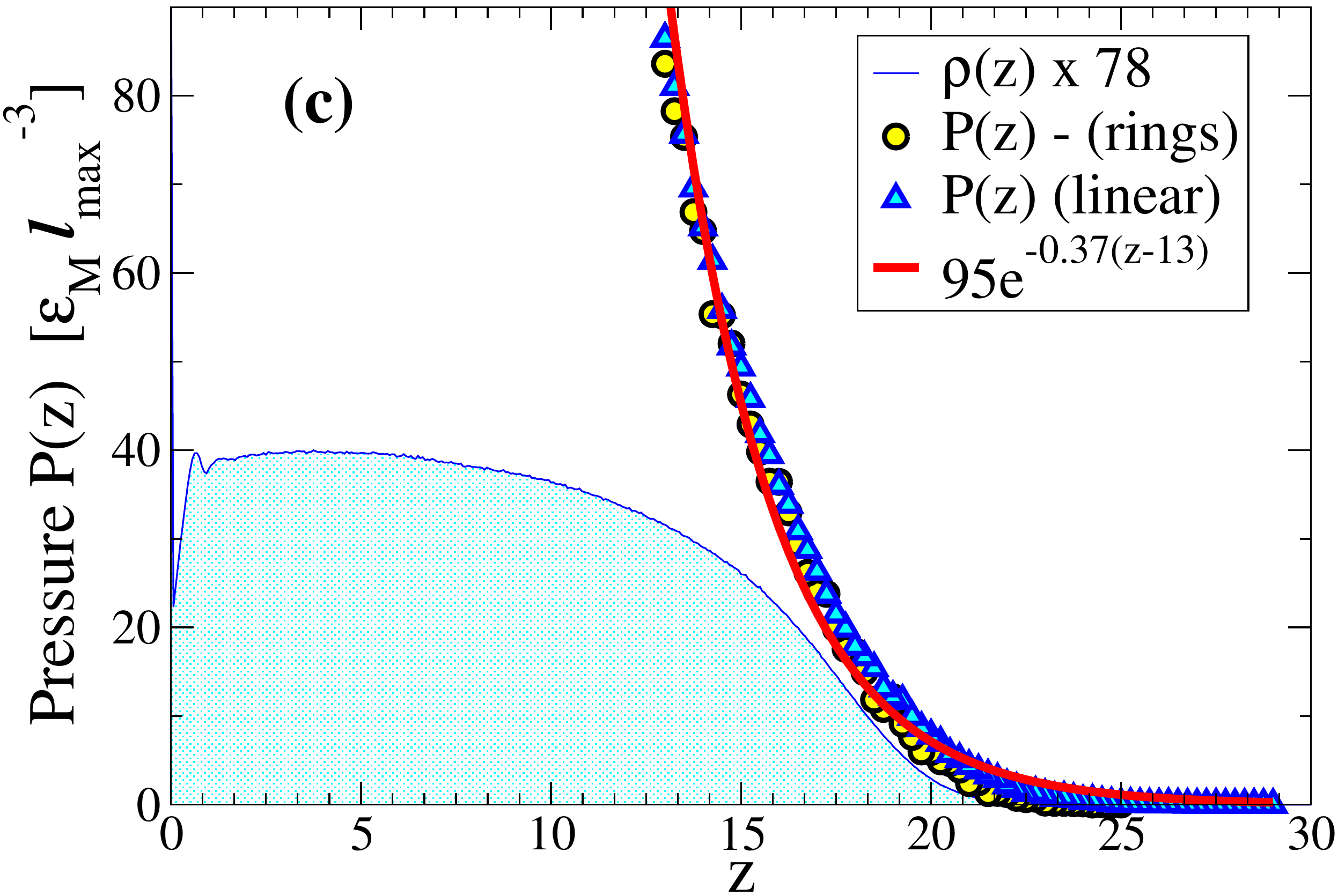}
\caption{(a) Density profiles along the $z$-direction for rings and linear
chains of different length (MD data). The density profiles scale linearly with
the chain length for linear chains respectively half the length for rings. In the inset
the curves collapse using this scaling relation. (b) Total monomer number
density $\rho(z)$, and end-monomers, $\rho_e(z)$  , respectively, middle-monomer
density $\rho_m(z)$ from MC data. Here $\sigma_g^R=0.0625$ (rings) and
$\sigma_g^L=0.125$ (linear chains). (c) Pressure, exerted
by the polymer brush of rings (circles), and linear chains (triangles) on a
piston at distance $z$ from the grafting plane - MC data for $\sigma_g^R=0.0625$
(rings) and $\sigma_g^L=0.125$ (linear chains). Here $N_R=128$, $N_L=64$. The density profile is
included for comparison.} \label{profiles}
\end{figure}
We always choose $L \times L$ grafting plane and periodic boundary conditions in x,y directions. 
Chains are grafted at a grafting
density $\sigma _{g,L}=0.125$ for the linear chains with the range
parameter $\sigma$ of the Weeks-Chandler-Andersen (WCA) potential that describes
the interaction between effective monomers in our MD simulations as unit of
length
\begin{equation} \label{eq2}
U_{WCA}(r) = 4 \varepsilon \Big[(\sigma/ r)^{12} - (\sigma/r)^6 + \frac{1}{4}
\Big] , \quad r < r_c=2^{1/6} \sigma \quad .
\end{equation}
$r$ being the distance between any pair of monomers (bonded or nonbonded ones).
$U_{WCA} (r \geq r_c) =0$, so the potential is purely repulsive. As is standard
\cite{Grest1999,Kremer1990} bonded monomers interact with the finitely
extensible nonlinear elastic (FENE) potential in addition,
\begin{equation} \label{eq3}
U_{\rm FENE} (r) = - 0.5 k R_0^2 \ln [1-(r/R_0)^2] \quad , \quad r < R_0  \quad ,
\end{equation}
where $U_{\rm FENE} (r > R_0) \equiv 0$. Parameters chosen are $\varepsilon=1$,
temperature $T=1.0\,\varepsilon/k_B$ and $k=30$, $R_0=1.5$. The mass of the
particles $m=1$ as well, so the MD time unit is $\tau=\sigma /
\sqrt{m/\varepsilon}=1$ as well. MD runs were carried out on GPUs using the
HooMD Blue $r$3574 code \cite{Anderson2008}. Note that due to the use of
GTX480 GPUs, a speedup factor of roughly 70 in comparison to a run on a single 
i7 CPU core for the present application is obtained. 
We use a standard 
dissipative particle
dynamics (DPD) thermostat \cite{Soddemann2003} with parameters $\gamma=0.5$,
$r_{\rm cut}=1.25\cdot 2^{1/6}$, and integration time step $\delta t =0.005$. The
grafting of the first monomer to the substrate site in the $z=0$ substrate plane
is realized by the same FENE potential as in Eq.~(\ref{eq3}). Note that the
linear dimension $L$ is adjusted such that we always have either
$\mathcal{N}=51200$ or $\mathcal{N}=102400$ effective monomers in the system,
for chain lengths $N_L=16,32,64,128,256$ and 512, respectively (recall $N_R=2
N_L$, $ \sigma_{g,R}=\sigma_{g,L}/2$ throughout). For the longest chains, an
equilibration time of $\tau_{\rm eq}\approx 2.1 \cdot 10^{8}$ was needed. 

MC simulations were done using a somewhat different off-lattice model that we describe
in the following. Of course, the motivation for choosing two different models is
to confirm the universality of our results, which should not depend on
irrelevant simulation details, but are rather generic.  We employ a model of
flexible polymer chains, each polymer consisting of $N_L$ or $N_R$ beads
connected by anharmonic springs. These springs are defined by the FENE
potential,  Eq.~(\ref{eq3}), albeit with $r = \ell - \ell _0$. Thus, $\ell$ is
the bond length, which can vary between $\ell _{min} < \ell < \ell_{max}$, and
has the equilibrium value $\ell _0 = 0.7$, while $\ell _{max}= \ell_0 +R_0$,
$\ell_{min} = \ell _0 - R_0$, and $R_0 = 0.3$. With these parameters $\ell
_{max} =1$ is the unit of the length. A short-range Morse potential is used to
describe non-bonded interactions,
\begin{equation} \label{Morse_pot}
U_M(r)=\varepsilon _M {\exp{[-2(r-r_{min}]} -2 \exp{[-x(r-r_{min})]}},
\end{equation}
with $\varepsilon _M =1$, $r_{min} =0.8$, $\alpha =24$. Since for this model
the $\theta$-temperature is $k_B\Theta \approx 0.62$, in the present study we
work at temperature $k_B\Theta = 1.0$, and are thus in the ``good solvent''
regime. The Monte Carlo procedure consists of local displacements of the monomers only, which are
accepted according to the usual Metropolis criterion. Typically, systems containing 32768 monomers are simulated about $6 \div 10 \times 10^6$
MCS after equilibration, whereby various quantities of interest are sampled.

\section{Results and Discussion}
\begin{figure}
\includegraphics{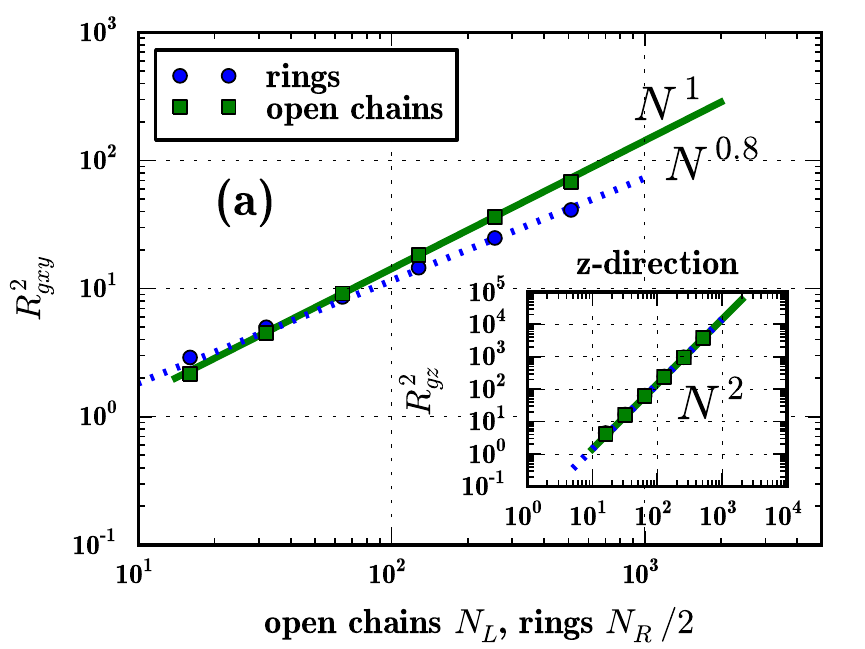}
\includegraphics[width=8.5cm]{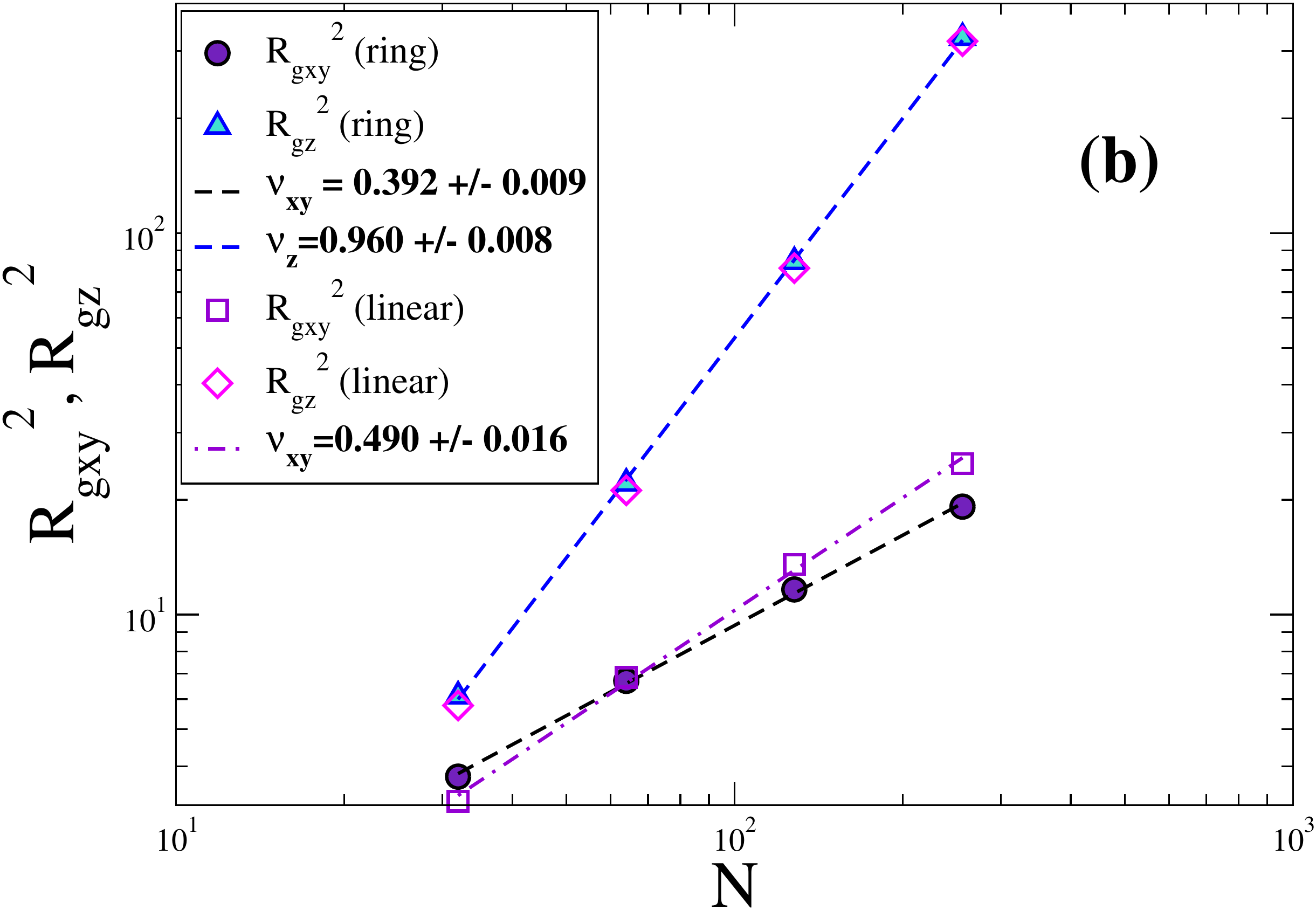}
\caption{(a) Squared radius of gyration (MD) parallel to the grafting plane on
double logarithmic scale $R_{gxy}^2=R_{gx}^2+ R_{gy}^2$ against chain
length $N_L$ (linear chains) and $N_R/2$ (rings), respectivly. Inset -
Scaling of the squared radius of gyration $R_{gz}^2$ perpendicular to the
grafting plane. The scaling prefactors differ by a factor of 4 between linear
chains and rings. (b) The same from MC data for $\sigma_g^R=0.0625$ (rings) and
$\sigma_g^L=0.125$ (linear chains). }
\label{Rg_scal}
\end{figure}
One of the central static properties of polymer brushes is the density profile
$\rho(z)$ of monomers in the $z$-direction perpendicular to the brush
(Fig.~\ref{profiles}a,~b). A striking finding is that for both models,
the density profiles of linear chains and rings are identical, within
statistical errors (which are larger for the MC calculation, due to the much
smaller statistical effort). The insert of Fig.~\ref{profiles}a shows the
profiles rescaling $z$ by $N_L$, to demonstrate that the data converge to the
strong stretching limit. For small $N_L$ there are oscillations for small $z$
and a ``finite size tail'' for large $z$: these ``corrections to scaling''
become negligible in the limit of very long chains. However, the variation of
$\rho(z)$ near the brush height $h$ (where $\rho(z)$ vanishes for large $N_L$)
is somewhat steeper than expected from the strong stretching limit of
self-consistent field theory \cite{Milner1988,Zhulina1991}, which would imply
that $\rho(z)/\rho(0)=1-(z/h)^2$: the latter result requires that the density
inside of the blobs does not exceed the semi-dilute regime \cite{DeGennes1979}.
However, one should note  that melt densities would be (in our units
\cite{Kremer1990}) of order $\rho \approx 1$ for the MD model and $\rho\approx 2$
for the MC model. It is clear that our choice of
parameters (leading to $\rho(0) \approx 0.3$, see Fig.~\ref{profiles}a)
corresponds to the regime of a rather concentrated solution rather than a 
semidilute solution (remember that Eq.~\ref{eq2} accounts for the effects 
of a good solvent only implicitly rather than explicitly, of course).

Fig.~\ref{profiles}b also includes the distribution of free chains ends (for the
linear chains) and of the mid-monomers with index $i=N_R/2$ for the rings (which
have $N_R=2N_L$ monomers, monomers being labeled consecutively along the ring,
$i=1$ being the grafted monomer). While according to the picture of the
``Alexander brush'' \cite{Alexander1977} $\rho_e(z) \propto \delta (z-h)$, all
end monomers need to be at the brush end, the actual distribution 
$\rho_e(z)$ shows that the ends (or mid-monomers in the case of rings, 
respectively) can be anywhere in the brush (although only few are near the 
grafting plane, since $\rho_e(z) \propto z$ for $z \ll h$ \cite{Milner1988}). 
It is interesting to
notice that in $\rho_e (z)$ there are slight but systematic differences between
linear chains and rings: for the latter the distributions are significantly more sharply
peaked, and the tail towards large $z$ is less pronounced. This is not
surprising, of course, since the mid-monomer is bound by two strands rather than
a single one. In Fig.~\ref{profiles}c we show the pressure, exerted by a brush or ring (circles), or
linear polymers (triangles) on a piston at height $z$ above the grafting plane. Evidently, the
pressure closely follows the monomer density profile at the brush tail, whereby one harldy detects
any difference between rings and linear chains.

A very interesting behavior results when we examine the linear dimensions of
individual chains, however (Fig.~\ref{Rg_scal}a,~\ref{Rg_scal}b). While the
$z$-components of the mean square gyration radii in $z$-directions $\langle
R_{gz} ^2 \rangle$ for linear chains and rings coincide precisely, the
transverse components differ significantly, indicating a different power law,
\begin{equation} \label{eq4}
\langle R^2_{gxy} \rangle \propto N \quad (\rm linear \, chains), \, \langle R^2_{gxy} \rangle \propto N^{0.8} \, ({\rm rings}) \quad.
\end{equation}

While the first of these equations is expected (Eq.~(\ref{eq1})), the second is
not. Surprisingly, our data coincide with the theoretical prediction for lattice
animals ($\nu=2/5)$ \cite{Cates1986} over a full decade in $N$, for both models.
For very small $N$, deviations occur, as expected, but indications for a
crossover to the ``crumpled globule'' exponent \cite{Grosberg1988} ($\nu=1/3)$
are not seen, unlike the case of bulk polymer melts
\cite{Muller2000,Muller2000a,Brown2001,Hur2006,Suzuki2008,Suzuki2009,
Vettorel2009,Vettorel2009a, Hur2011,Halverson2010, Halverson2010a}. Of course,
while it is conceivable that another crossover could occur for much
larger ring lengths, it is not really clear that one should expect this: after
all, the $z$-components of the grafted chains in the brush are strongly
stretched, $\langle R^2_{gz} \rangle \propto N^2$, while in a ring polymer melt
all components $R^2_{gx}$, $R^2_{gy}$ and $R^2_{gz}$ scale in the same way.
Due to this strong anisotropy, it remains unclear if the transverse components
$R^2_{gxy}$ of the chains in a ring polymer brush should exhibit the same
scaling as chains in a ring polymer melt.
\begin{figure}
\includegraphics{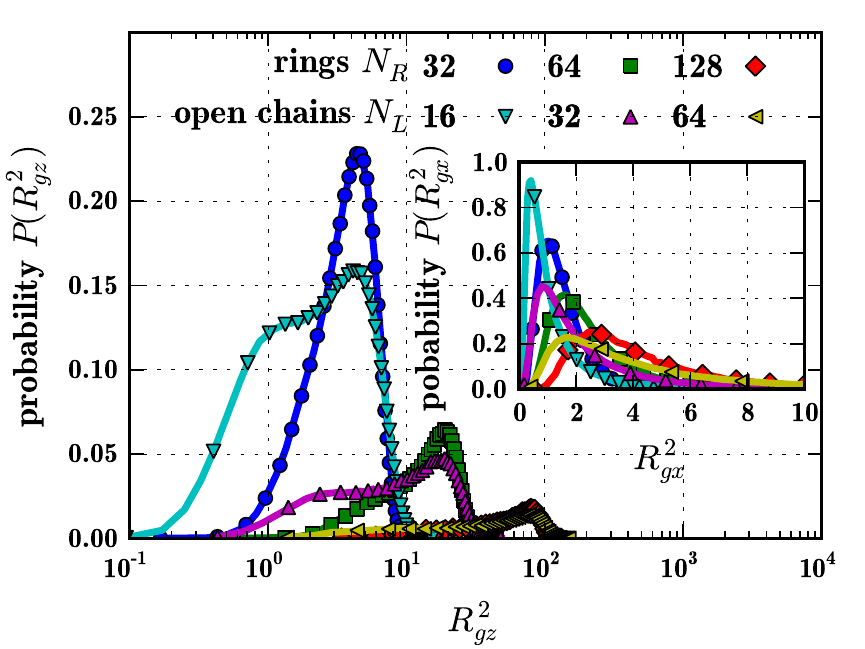}
\caption{Normalized probability distributions of  $R_{gz}^2$ for rings and
linear chains of different length $N$. The shoulders in the probability distribution
of the open chains towards lower values of $R_{gz}^2$ indicate that this
distribution is non-gaussian -  there is a higher probability of finding compact
configurations of linear chains as compared to rings of double length. {\it
Inset:} Normalized probability distribution of the squared radius of gyration in
x-direction parallel to the grafting plane $R_{gx}^2$.}
\label{Rg_PDF}
\end{figure}
The fact that the mean-square gyration radii in $z$-direction, $\langle
R^2_{gz} \rangle$, for linear chains and rings (under equivalent conditions,
$N_R=2N_L$, $\sigma^R_g= \sigma_g^L/2$) coincide, while the transverse
components are so different in both cases, should not be mistaken as a proof
that $z$-components and transverse components are decoupled, however. When one
compares the full probability distributions $P(R^2_{gz})$ for rings and linear
chains, one does see characteristic differences: for very small $R^2
_{gz}$, the distribution in the case of rings has much more weight than it has
for linear chains (Fig.~\ref{Rg_PDF}). Of course, this finding does not
contradict Figs.~\ref{Rg_scal}a,~\ref{Rg_scal}b, since the region of very small
$R^2_{gz}$ makes only negligibly small contributions to $\langle R^2_{gz}
\rangle$.
\begin{figure*}
\includegraphics{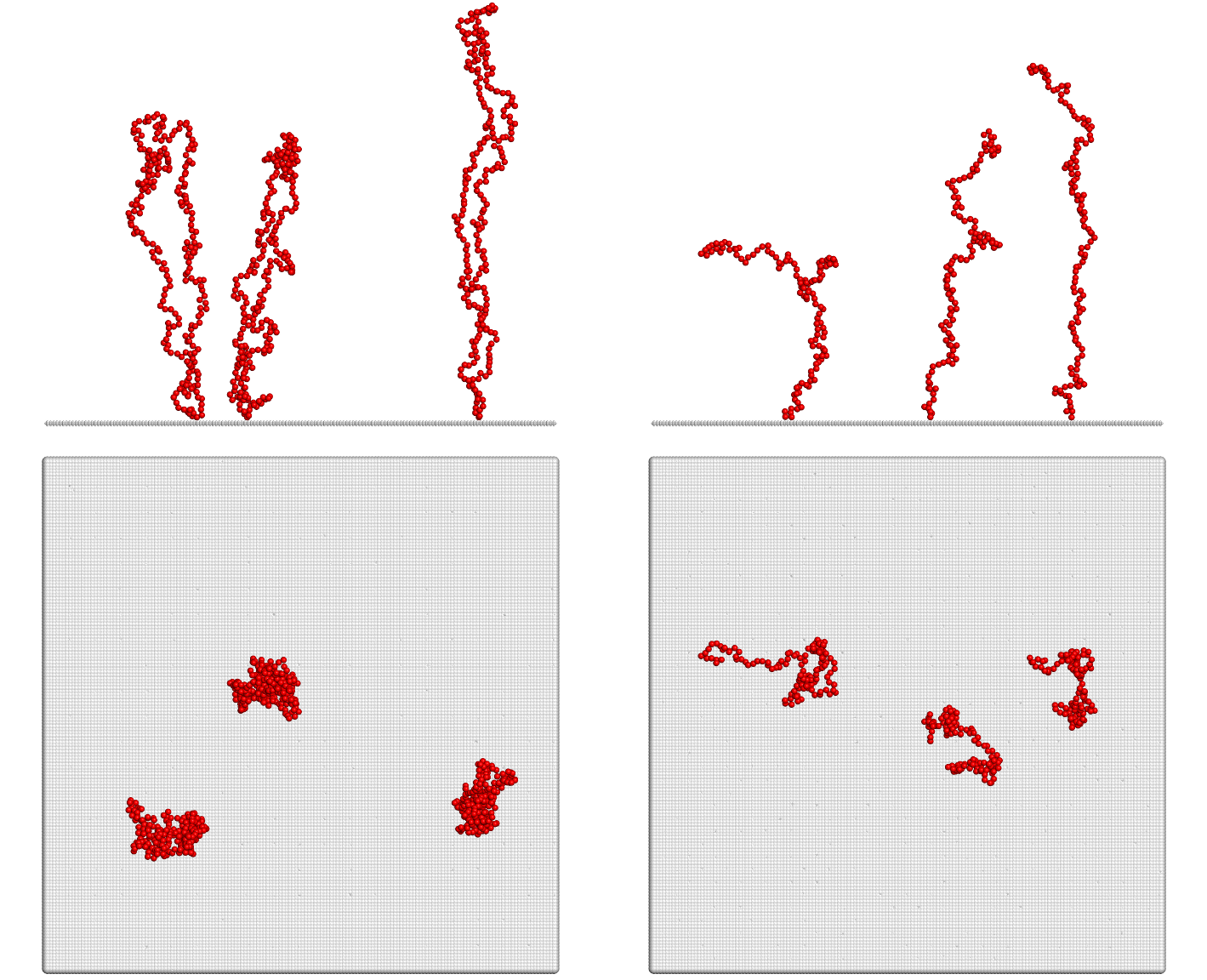}
\caption{Configurations from Molecular dynamics simulations to illustrate the
conformational differences between loops and open chains: 3 out of 400 loops of ring
length $N_R = 256$ ({\it left}) and 3 out of 800 open chains of chain length $N_L = 128$ ({\it
right}) are displayed. The rings and the open chains are grafted on a $80\times 80$ grafting plane. The overall
number of monomers is 102400 per system.}
\label{snapshots}
\end{figure*}
\section{Concluding summary} In this letter, we consider a new class of polymer
brushes, formed by densely grafted ring polymers to a flat substrate surface. We demonstrate that the
transverse configuration of such grafted rings differs substantially from corresponding linear chains
(Fig.~\ref{Rg_scal}) although collective properties (such as the average density profiles of the monomers
(Figs.~\ref{profiles}a,b) or the pressure exerted by the brush on a compressing wall (\ref{profiles}c))
basically do not differ at all. An intriguing question is the interpretation of our findings in
terms of geometric properties of chain versus ring conformations (Fig~\ref{snapshots} shows some
illustrative snapshots). In the corresponding problem of non-catenated rings forming dense melts,
one associates a $R_{g} \sim N_R^{2/5}$ variation with lattice animal like conformations (while for
very large ring-lengths $N_R$ one rather finds \cite{Halverson2010} crumpled globules, leading to
$R_g \sim N_{R}^{1/3}$). Obviously, the projections of monomer positions into the xy-plane for rings
(Fig.~\ref{snapshots}) are fairly compact, while the corresponding projections for linear chains are
random walk-like.

An intriguing question is what phenomena are affected by these different conformations of rings
(rather than linear chains) forming a polymer brush. Presumably, for instance, the dynamics of free
(linear) chains from a solution or melt flowing past a ring polymer brush will be very different.
The interaction with nanoparticle inclusions, that involves the deformation of chains (or
rings) surrounding the nanoparticle would be rather different. All such questions remain to be 
explored also.

Of course, it would be very nice if guidance from analytical theory were
available to understand the differences between ring polymer brushes and
ordinary brushes formed from linear polymers, as we have found in our studies.
However, none of the well-established approaches (self-consistent field theory
\cite{Milner1988,Zhulina1991,Netz1998}, density functional-type theories
\cite{Szleifer1996,Egorov2008}, etc) can take the topological differences between
catenated and non-catenated rings into account: none of the theories could hence
predict any of the differences that we have found here. Also experimental studies of this problem
(ideally one migth consider neutron scattering from deuterated rings in a protonated ring polymer
brush) are a great challenge.

\acknowledgments
We thank the Deutsche Forschungsgemeinschaft (DFG) for partial support under grant number SFB
625/A17 (D.R.) and BI 314/23 (A.M.) and are grateful to K. Kremer for sending preprints of Refs.
39,40 and stimulating discussions. Computing time on the GPU cluster of the Center for Computational
Sciences Mainz at the ZDV Mainz are gratefully acknowledged, too.

\bibliography{literatur5}

\begin{thebibliography}{57}
\expandafter\ifx\csname natexlab\endcsname\relax\def\natexlab#1{#1}\fi
\expandafter\ifx\csname bibnamefont\endcsname\relax
  \def\bibnamefont#1{#1}\fi
\expandafter\ifx\csname bibfnamefont\endcsname\relax
  \def\bibfnamefont#1{#1}\fi
\expandafter\ifx\csname citenamefont\endcsname\relax
  \def\citenamefont#1{#1}\fi
\expandafter\ifx\csname url\endcsname\relax
  \def\url#1{\texttt{#1}}\fi
\expandafter\ifx\csname urlprefix\endcsname\relax\def\urlprefix{URL }\fi
\providecommand{\bibinfo}[2]{#2}
\providecommand{\eprint}[2][]{\url{#2}}

\bibitem[{\citenamefont{Halperin et~al.}(1992)\citenamefont{Halperin, Tirrell,
  and Lodge}}]{Halperin1992}
\bibinfo{author}{\bibfnamefont{A.}~\bibnamefont{Halperin}},
  \bibinfo{author}{\bibfnamefont{M.}~\bibnamefont{Tirrell}}, \bibnamefont{and}
  \bibinfo{author}{\bibfnamefont{T.}~\bibnamefont{Lodge}},
  \bibinfo{journal}{Adv. Polym. Sci.} \textbf{\bibinfo{volume}{100}},
  \bibinfo{pages}{31} (\bibinfo{year}{1992}).

\bibitem[{\citenamefont{Klein}(1996)}]{Klein1996}
\bibinfo{author}{\bibfnamefont{J.}~\bibnamefont{Klein}},
  \bibinfo{journal}{Annu. Rev. Mater. Sci.} \textbf{\bibinfo{volume}{26}},
  \bibinfo{pages}{581} (\bibinfo{year}{1996}).

\bibitem[{\citenamefont{Szleifer and Carignano}(1996)}]{Szleifer1996}
\bibinfo{author}{\bibfnamefont{I.}~\bibnamefont{Szleifer}} \bibnamefont{and}
  \bibinfo{author}{\bibfnamefont{M.~A.} \bibnamefont{Carignano}},
  \bibinfo{journal}{Adv. Chem. Phys.} \textbf{\bibinfo{volume}{94}},
  \bibinfo{pages}{165} (\bibinfo{year}{1996}).

\bibitem[{\citenamefont{Grest}(1999)}]{Grest1999}
\bibinfo{author}{\bibfnamefont{G.~S.} \bibnamefont{Grest}},
  \bibinfo{journal}{Adv. Polym. Sci.} \textbf{\bibinfo{volume}{138}},
  \bibinfo{pages}{149} (\bibinfo{year}{1999}).

\bibitem[{\citenamefont{Advincula et~al.}(2004)\citenamefont{Advincula,
  Brittain, Caster, and R\"{u}he}}]{Advincula2004}
\bibinfo{editor}{\bibfnamefont{R.~C.} \bibnamefont{Advincula}},
  \bibinfo{editor}{\bibfnamefont{W.~J.} \bibnamefont{Brittain}},
  \bibinfo{editor}{\bibfnamefont{K.~C.} \bibnamefont{Caster}},
  \bibnamefont{and} \bibinfo{editor}{\bibfnamefont{J.}~\bibnamefont{R\"{u}he}},
  eds., \emph{\bibinfo{title}{{Polymer brushes}}}
  (\bibinfo{publisher}{Wiley-VCH}, \bibinfo{address}{Weinheim},
  \bibinfo{year}{2004}).

\bibitem[{\citenamefont{Binder et~al.}(2011)\citenamefont{Binder, Kreer, and
  Milchev}}]{Binder2011}
\bibinfo{author}{\bibfnamefont{K.}~\bibnamefont{Binder}},
  \bibinfo{author}{\bibfnamefont{T.}~\bibnamefont{Kreer}}, \bibnamefont{and}
  \bibinfo{author}{\bibfnamefont{A.}~\bibnamefont{Milchev}},
  \bibinfo{journal}{Soft Matter} p. \bibinfo{pages}{in press}
  (\bibinfo{year}{2011}).

\bibitem[{\citenamefont{Napper}(1983)}]{Napper1983}
\bibinfo{author}{\bibfnamefont{D.}~\bibnamefont{Napper}},
  \emph{\bibinfo{title}{{Polymeric Stabilization of Colloidal Dispersions}}}
  (\bibinfo{publisher}{Academic}, \bibinfo{address}{London},
  \bibinfo{year}{1983}).

\bibitem[{\citenamefont{Klein}(2009)}]{Klein2009}
\bibinfo{author}{\bibfnamefont{J.}~\bibnamefont{Klein}},
  \bibinfo{journal}{Science} \textbf{\bibinfo{volume}{323}},
  \bibinfo{pages}{47} (\bibinfo{year}{2009}).

\bibitem[{\citenamefont{Merlitz et~al.}(2009)\citenamefont{Merlitz, He, Wu, and
  Sommer}}]{Merlitz2009}
\bibinfo{author}{\bibfnamefont{H.}~\bibnamefont{Merlitz}},
  \bibinfo{author}{\bibfnamefont{G.-L.} \bibnamefont{He}},
  \bibinfo{author}{\bibfnamefont{C.-X.} \bibnamefont{Wu}}, \bibnamefont{and}
  \bibinfo{author}{\bibfnamefont{J.-U.} \bibnamefont{Sommer}},
  \bibinfo{journal}{Phys. Rev. Lett.} \textbf{\bibinfo{volume}{102}},
  \bibinfo{pages}{115702} (\bibinfo{year}{2009}).

\bibitem[{\citenamefont{Storm}(1995)}]{Storm1995}
\bibinfo{author}{\bibfnamefont{G.}~\bibnamefont{Storm}}, \bibinfo{journal}{Adv.
  Drug. Deliv. Rev.} \textbf{\bibinfo{volume}{17}}, \bibinfo{pages}{31}
  (\bibinfo{year}{1995}).

\bibitem[{\citenamefont{Wang et~al.}(2007)\citenamefont{Wang, Xu, and
  Chen}}]{Wang2007}
\bibinfo{author}{\bibfnamefont{A.-J.} \bibnamefont{Wang}},
  \bibinfo{author}{\bibfnamefont{J.-J.} \bibnamefont{Xu}}, \bibnamefont{and}
  \bibinfo{author}{\bibfnamefont{H.-Y.} \bibnamefont{Chen}},
  \bibinfo{journal}{J. Chromatogr. A} \textbf{\bibinfo{volume}{1147}},
  \bibinfo{pages}{120} (\bibinfo{year}{2007}).

\bibitem[{\citenamefont{Alexander}(1977)}]{Alexander1977}
\bibinfo{author}{\bibfnamefont{S.}~\bibnamefont{Alexander}},
  \bibinfo{journal}{J. Phys.} \textbf{\bibinfo{volume}{38}},
  \bibinfo{pages}{977} (\bibinfo{year}{1977}).

\bibitem[{\citenamefont{de~Gennes}(1980)}]{deGennes1980}
\bibinfo{author}{\bibfnamefont{P.~G.} \bibnamefont{de~Gennes}},
  \bibinfo{journal}{Macromolecules} \textbf{\bibinfo{volume}{13}},
  \bibinfo{pages}{1069} (\bibinfo{year}{1980}).

\bibitem[{\citenamefont{Milner et~al.}(1988)\citenamefont{Milner, Witten, and
  Cates}}]{Milner1988}
\bibinfo{author}{\bibfnamefont{S.~T.} \bibnamefont{Milner}},
  \bibinfo{author}{\bibfnamefont{T.~a.} \bibnamefont{Witten}},
  \bibnamefont{and} \bibinfo{author}{\bibfnamefont{M.~E.} \bibnamefont{Cates}},
  \bibinfo{journal}{Macromolecules} \textbf{\bibinfo{volume}{21}},
  \bibinfo{pages}{2610} (\bibinfo{year}{1988}).

\bibitem[{\citenamefont{Zhulina et~al.}(1991)\citenamefont{Zhulina, Borisov,
  Pryamitsyn, and Birshtein}}]{Zhulina1991}
\bibinfo{author}{\bibfnamefont{E.~B.} \bibnamefont{Zhulina}},
  \bibinfo{author}{\bibfnamefont{O.~V.} \bibnamefont{Borisov}},
  \bibinfo{author}{\bibfnamefont{V.~a.} \bibnamefont{Pryamitsyn}},
  \bibnamefont{and} \bibinfo{author}{\bibfnamefont{T.~M.}
  \bibnamefont{Birshtein}}, \bibinfo{journal}{Macromolecules}
  \textbf{\bibinfo{volume}{24}}, \bibinfo{pages}{140} (\bibinfo{year}{1991}).

\bibitem[{\citenamefont{Netz and Schick}(1998)}]{Netz1998}
\bibinfo{author}{\bibfnamefont{R.}~\bibnamefont{Netz}} \bibnamefont{and}
  \bibinfo{author}{\bibfnamefont{M.}~\bibnamefont{Schick}},
  \bibinfo{journal}{Macromolecules} \textbf{\bibinfo{volume}{31}},
  \bibinfo{pages}{5105} (\bibinfo{year}{1998}).

\bibitem[{\citenamefont{Kreer et~al.}(2004)\citenamefont{Kreer, Metzger,
  M\"{u}ller, Binder, and Baschnagel}}]{Kreer2004}
\bibinfo{author}{\bibfnamefont{T.}~\bibnamefont{Kreer}},
  \bibinfo{author}{\bibfnamefont{S.}~\bibnamefont{Metzger}},
  \bibinfo{author}{\bibfnamefont{M.}~\bibnamefont{M\"{u}ller}},
  \bibinfo{author}{\bibfnamefont{K.}~\bibnamefont{Binder}}, \bibnamefont{and}
  \bibinfo{author}{\bibfnamefont{J.}~\bibnamefont{Baschnagel}},
  \bibinfo{journal}{J. Chem. Phys.} \textbf{\bibinfo{volume}{120}},
  \bibinfo{pages}{4012} (\bibinfo{year}{2004}).

\bibitem[{\citenamefont{Galuschko et~al.}(2010)\citenamefont{Galuschko, Spirin,
  Kreer, Johner, Pastorino, Wittmer, and Baschnagel}}]{Galuschko2010}
\bibinfo{author}{\bibfnamefont{A.}~\bibnamefont{Galuschko}},
  \bibinfo{author}{\bibfnamefont{L.}~\bibnamefont{Spirin}},
  \bibinfo{author}{\bibfnamefont{T.}~\bibnamefont{Kreer}},
  \bibinfo{author}{\bibfnamefont{A.}~\bibnamefont{Johner}},
  \bibinfo{author}{\bibfnamefont{C.}~\bibnamefont{Pastorino}},
  \bibinfo{author}{\bibfnamefont{J.}~\bibnamefont{Wittmer}}, \bibnamefont{and}
  \bibinfo{author}{\bibfnamefont{J.}~\bibnamefont{Baschnagel}},
  \bibinfo{journal}{Langmuir} \textbf{\bibinfo{volume}{26}},
  \bibinfo{pages}{6418} (\bibinfo{year}{2010}).

\bibitem[{\citenamefont{Yin et~al.}(2007)\citenamefont{Yin, Bedrov, Smith, and
  Kilbey}}]{Yin2007}
\bibinfo{author}{\bibfnamefont{F.}~\bibnamefont{Yin}},
  \bibinfo{author}{\bibfnamefont{D.}~\bibnamefont{Bedrov}},
  \bibinfo{author}{\bibfnamefont{G.~D.} \bibnamefont{Smith}}, \bibnamefont{and}
  \bibinfo{author}{\bibfnamefont{S.~M.} \bibnamefont{Kilbey}},
  \bibinfo{journal}{J. Chem. Phys.} \textbf{\bibinfo{volume}{127}},
  \bibinfo{pages}{084910} (\bibinfo{year}{2007}).

\bibitem[{\citenamefont{Roovers}(1988)}]{Roovers1988}
\bibinfo{author}{\bibfnamefont{J.}~\bibnamefont{Roovers}},
  \bibinfo{journal}{Macromolecules} \textbf{\bibinfo{volume}{21}},
  \bibinfo{pages}{1517} (\bibinfo{year}{1988}).

\bibitem[{\citenamefont{McKenna et~al.}(1989)\citenamefont{McKenna, Hostetter,
  Hadjichristidis, Fetters, and Plazek}}]{McKenna1989}
\bibinfo{author}{\bibfnamefont{G.~B.} \bibnamefont{McKenna}},
  \bibinfo{author}{\bibfnamefont{B.~J.} \bibnamefont{Hostetter}},
  \bibinfo{author}{\bibfnamefont{N.}~\bibnamefont{Hadjichristidis}},
  \bibinfo{author}{\bibfnamefont{L.~J.} \bibnamefont{Fetters}},
  \bibnamefont{and} \bibinfo{author}{\bibfnamefont{D.~J.}
  \bibnamefont{Plazek}}, \bibinfo{journal}{Macromolecules}
  \textbf{\bibinfo{volume}{22}}, \bibinfo{pages}{1834} (\bibinfo{year}{1989}).

\bibitem[{\citenamefont{Gagliardi et~al.}(2005)\citenamefont{Gagliardi,
  Arrighi, Ferguson, Dagger, Semlyen, and Higgins}}]{Gagliardi2005}
\bibinfo{author}{\bibfnamefont{S.}~\bibnamefont{Gagliardi}},
  \bibinfo{author}{\bibfnamefont{V.}~\bibnamefont{Arrighi}},
  \bibinfo{author}{\bibfnamefont{R.}~\bibnamefont{Ferguson}},
  \bibinfo{author}{\bibfnamefont{a.~C.} \bibnamefont{Dagger}},
  \bibinfo{author}{\bibfnamefont{J.~a.} \bibnamefont{Semlyen}},
  \bibnamefont{and} \bibinfo{author}{\bibfnamefont{J.~S.}
  \bibnamefont{Higgins}}, \bibinfo{journal}{J. Chem. Phys.}
  \textbf{\bibinfo{volume}{122}}, \bibinfo{pages}{064904}
  (\bibinfo{year}{2005}).

\bibitem[{\citenamefont{Kawaguchi et~al.}(2006)\citenamefont{Kawaguchi,
  Masuoka, Takano, Tanaka, Nagamura, Torikai, Dalgliesh, Langridge, and
  Matsushita}}]{Kawaguchi2006}
\bibinfo{author}{\bibfnamefont{D.}~\bibnamefont{Kawaguchi}},
  \bibinfo{author}{\bibfnamefont{K.}~\bibnamefont{Masuoka}},
  \bibinfo{author}{\bibfnamefont{A.}~\bibnamefont{Takano}},
  \bibinfo{author}{\bibfnamefont{K.}~\bibnamefont{Tanaka}},
  \bibinfo{author}{\bibfnamefont{T.}~\bibnamefont{Nagamura}},
  \bibinfo{author}{\bibfnamefont{N.}~\bibnamefont{Torikai}},
  \bibinfo{author}{\bibfnamefont{R.~M.} \bibnamefont{Dalgliesh}},
  \bibinfo{author}{\bibfnamefont{S.}~\bibnamefont{Langridge}},
  \bibnamefont{and}
  \bibinfo{author}{\bibfnamefont{Y.}~\bibnamefont{Matsushita}},
  \bibinfo{journal}{Macromolecules} \textbf{\bibinfo{volume}{39}},
  \bibinfo{pages}{5180} (\bibinfo{year}{2006}).

\bibitem[{\citenamefont{Kapnistos et~al.}(2008)\citenamefont{Kapnistos, Lang,
  Vlassopoulos, Pyckhout-Hintzen, Richter, Cho, Chang, and
  Rubinstein}}]{Kapnistos2008}
\bibinfo{author}{\bibfnamefont{M.}~\bibnamefont{Kapnistos}},
  \bibinfo{author}{\bibfnamefont{M.}~\bibnamefont{Lang}},
  \bibinfo{author}{\bibfnamefont{D.}~\bibnamefont{Vlassopoulos}},
  \bibinfo{author}{\bibfnamefont{W.}~\bibnamefont{Pyckhout-Hintzen}},
  \bibinfo{author}{\bibfnamefont{D.}~\bibnamefont{Richter}},
  \bibinfo{author}{\bibfnamefont{D.}~\bibnamefont{Cho}},
  \bibinfo{author}{\bibfnamefont{T.}~\bibnamefont{Chang}}, \bibnamefont{and}
  \bibinfo{author}{\bibfnamefont{M.}~\bibnamefont{Rubinstein}},
  \bibinfo{journal}{Nature Mater.} \textbf{\bibinfo{volume}{7}},
  \bibinfo{pages}{997} (\bibinfo{year}{2008}).

\bibitem[{\citenamefont{Nam et~al.}(2009)\citenamefont{Nam, Leisen, Breedveld,
  and Beckham}}]{Nam2009}
\bibinfo{author}{\bibfnamefont{S.}~\bibnamefont{Nam}},
  \bibinfo{author}{\bibfnamefont{J.}~\bibnamefont{Leisen}},
  \bibinfo{author}{\bibfnamefont{V.}~\bibnamefont{Breedveld}},
  \bibnamefont{and} \bibinfo{author}{\bibfnamefont{H.~W.}
  \bibnamefont{Beckham}}, \bibinfo{journal}{Macromolecules}
  \textbf{\bibinfo{volume}{42}}, \bibinfo{pages}{3121} (\bibinfo{year}{2009}).

\bibitem[{\citenamefont{Cates and Deutsch}(1986)}]{Cates1986}
\bibinfo{author}{\bibfnamefont{M.}~\bibnamefont{Cates}} \bibnamefont{and}
  \bibinfo{author}{\bibfnamefont{J.}~\bibnamefont{Deutsch}},
  \bibinfo{journal}{J. Phys.} \textbf{\bibinfo{volume}{47}},
  \bibinfo{pages}{2121} (\bibinfo{year}{1986}).

\bibitem[{\citenamefont{M\"{u}ller et~al.}(1996)\citenamefont{M\"{u}ller,
  Wittmer, and Cates}}]{Muller1996}
\bibinfo{author}{\bibfnamefont{M.}~\bibnamefont{M\"{u}ller}},
  \bibinfo{author}{\bibfnamefont{J.}~\bibnamefont{Wittmer}}, \bibnamefont{and}
  \bibinfo{author}{\bibfnamefont{M.}~\bibnamefont{Cates}},
  \bibinfo{journal}{Phys. Rev. E.} \textbf{\bibinfo{volume}{53}},
  \bibinfo{pages}{5063} (\bibinfo{year}{1996}).

\bibitem[{\citenamefont{Brown and Szamel}(1998{\natexlab{a}})}]{Brown1998}
\bibinfo{author}{\bibfnamefont{S.}~\bibnamefont{Brown}} \bibnamefont{and}
  \bibinfo{author}{\bibfnamefont{G.}~\bibnamefont{Szamel}},
  \bibinfo{journal}{J. Chem. Phys.} \textbf{\bibinfo{volume}{108}},
  \bibinfo{pages}{4705} (\bibinfo{year}{1998}{\natexlab{a}}).

\bibitem[{\citenamefont{Brown and Szamel}(1998{\natexlab{b}})}]{Brown1998a}
\bibinfo{author}{\bibfnamefont{S.}~\bibnamefont{Brown}} \bibnamefont{and}
  \bibinfo{author}{\bibfnamefont{G.}~\bibnamefont{Szamel}},
  \bibinfo{journal}{J. Chem. Phys.} \textbf{\bibinfo{volume}{109}},
  \bibinfo{pages}{6184} (\bibinfo{year}{1998}{\natexlab{b}}).

\bibitem[{\citenamefont{M\"{u}ller
  et~al.}(2000{\natexlab{a}})\citenamefont{M\"{u}ller, Wittmer, and
  Barrat}}]{Muller2000a}
\bibinfo{author}{\bibfnamefont{M.}~\bibnamefont{M\"{u}ller}},
  \bibinfo{author}{\bibfnamefont{J.~P.} \bibnamefont{Wittmer}},
  \bibnamefont{and} \bibinfo{author}{\bibfnamefont{J.-L.}
  \bibnamefont{Barrat}}, \bibinfo{journal}{Europhys. Lett.}
  \textbf{\bibinfo{volume}{52}}, \bibinfo{pages}{406}
  (\bibinfo{year}{2000}{\natexlab{a}}).

\bibitem[{\citenamefont{M\"{u}ller
  et~al.}(2000{\natexlab{b}})\citenamefont{M\"{u}ller, Wittmer, and
  Cates}}]{Muller2000}
\bibinfo{author}{\bibfnamefont{M.}~\bibnamefont{M\"{u}ller}},
  \bibinfo{author}{\bibfnamefont{J.}~\bibnamefont{Wittmer}}, \bibnamefont{and}
  \bibinfo{author}{\bibfnamefont{M.}~\bibnamefont{Cates}},
  \bibinfo{journal}{Phys. Rev. E.} \textbf{\bibinfo{volume}{61}},
  \bibinfo{pages}{4078} (\bibinfo{year}{2000}{\natexlab{b}}).

\bibitem[{\citenamefont{Brown et~al.}(2001)\citenamefont{Brown, Lenczycki, and
  Szamel}}]{Brown2001}
\bibinfo{author}{\bibfnamefont{S.}~\bibnamefont{Brown}},
  \bibinfo{author}{\bibfnamefont{T.}~\bibnamefont{Lenczycki}},
  \bibnamefont{and} \bibinfo{author}{\bibfnamefont{G.}~\bibnamefont{Szamel}},
  \bibinfo{journal}{Phys. Rev. E.} \textbf{\bibinfo{volume}{63}},
  \bibinfo{pages}{3} (\bibinfo{year}{2001}).

\bibitem[{\citenamefont{Hur et~al.}(2006)\citenamefont{Hur, Winkler, and
  Yoon}}]{Hur2006}
\bibinfo{author}{\bibfnamefont{K.}~\bibnamefont{Hur}},
  \bibinfo{author}{\bibfnamefont{R.~G.} \bibnamefont{Winkler}},
  \bibnamefont{and} \bibinfo{author}{\bibfnamefont{D.~Y.} \bibnamefont{Yoon}},
  \bibinfo{journal}{Macromolecules} \textbf{\bibinfo{volume}{39}},
  \bibinfo{pages}{3975} (\bibinfo{year}{2006}).

\bibitem[{\citenamefont{Suzuki et~al.}(2008)\citenamefont{Suzuki, Takano, and
  Matsushita}}]{Suzuki2008}
\bibinfo{author}{\bibfnamefont{J.}~\bibnamefont{Suzuki}},
  \bibinfo{author}{\bibfnamefont{A.}~\bibnamefont{Takano}}, \bibnamefont{and}
  \bibinfo{author}{\bibfnamefont{Y.}~\bibnamefont{Matsushita}},
  \bibinfo{journal}{J. Chem. Phys.} \textbf{\bibinfo{volume}{129}},
  \bibinfo{pages}{034903} (\bibinfo{year}{2008}).

\bibitem[{\citenamefont{Suzuki et~al.}(2009)\citenamefont{Suzuki, Takano,
  Deguchi, and Matsushita}}]{Suzuki2009}
\bibinfo{author}{\bibfnamefont{J.}~\bibnamefont{Suzuki}},
  \bibinfo{author}{\bibfnamefont{A.}~\bibnamefont{Takano}},
  \bibinfo{author}{\bibfnamefont{T.}~\bibnamefont{Deguchi}}, \bibnamefont{and}
  \bibinfo{author}{\bibfnamefont{Y.}~\bibnamefont{Matsushita}},
  \bibinfo{journal}{J. Chem. Phys.} \textbf{\bibinfo{volume}{131}},
  \bibinfo{pages}{144902} (\bibinfo{year}{2009}).

\bibitem[{\citenamefont{Vettorel
  et~al.}(2009{\natexlab{a}})\citenamefont{Vettorel, Reigh, Yoon, and
  Kremer}}]{Vettorel2009a}
\bibinfo{author}{\bibfnamefont{T.}~\bibnamefont{Vettorel}},
  \bibinfo{author}{\bibfnamefont{S.~Y.} \bibnamefont{Reigh}},
  \bibinfo{author}{\bibfnamefont{D.~Y.} \bibnamefont{Yoon}}, \bibnamefont{and}
  \bibinfo{author}{\bibfnamefont{K.}~\bibnamefont{Kremer}},
  \bibinfo{journal}{Macromol. Rapid Commun.} \textbf{\bibinfo{volume}{30}},
  \bibinfo{pages}{345} (\bibinfo{year}{2009}{\natexlab{a}}).

\bibitem[{\citenamefont{Vettorel
  et~al.}(2009{\natexlab{b}})\citenamefont{Vettorel, Grosberg, and
  Kremer}}]{Vettorel2009}
\bibinfo{author}{\bibfnamefont{T.}~\bibnamefont{Vettorel}},
  \bibinfo{author}{\bibfnamefont{A.~Y.} \bibnamefont{Grosberg}},
  \bibnamefont{and} \bibinfo{author}{\bibfnamefont{K.}~\bibnamefont{Kremer}},
  \bibinfo{journal}{Phys. Biol.} \textbf{\bibinfo{volume}{6}},
  \bibinfo{pages}{025013} (\bibinfo{year}{2009}{\natexlab{b}}).

\bibitem[{\citenamefont{Hur et~al.}(2011)\citenamefont{Hur, Jeong, Winkler,
  Lacevic, Gee, and Yoon}}]{Hur2011}
\bibinfo{author}{\bibfnamefont{K.}~\bibnamefont{Hur}},
  \bibinfo{author}{\bibfnamefont{C.}~\bibnamefont{Jeong}},
  \bibinfo{author}{\bibfnamefont{R.~G.} \bibnamefont{Winkler}},
  \bibinfo{author}{\bibfnamefont{N.}~\bibnamefont{Lacevic}},
  \bibinfo{author}{\bibfnamefont{R.~H.} \bibnamefont{Gee}}, \bibnamefont{and}
  \bibinfo{author}{\bibfnamefont{D.~Y.} \bibnamefont{Yoon}},
  \bibinfo{journal}{Macromolecules} \textbf{\bibinfo{volume}{44}},
  \bibinfo{pages}{2311} (\bibinfo{year}{2011}).

\bibitem[{\citenamefont{Halverson
  et~al.}(2010{\natexlab{a}})\citenamefont{Halverson, Lee, Grest, Grosberg, and
  Kremer}}]{Halverson2010}
\bibinfo{author}{\bibfnamefont{J.~D.} \bibnamefont{Halverson}},
  \bibinfo{author}{\bibfnamefont{W.~B.} \bibnamefont{Lee}},
  \bibinfo{author}{\bibfnamefont{G.~S.} \bibnamefont{Grest}},
  \bibinfo{author}{\bibfnamefont{A.~Y.} \bibnamefont{Grosberg}},
  \bibnamefont{and} \bibinfo{author}{\bibfnamefont{K.}~\bibnamefont{Kremer}}
  (\bibinfo{year}{2010}{\natexlab{a}}).

\bibitem[{\citenamefont{Halverson
  et~al.}(2010{\natexlab{b}})\citenamefont{Halverson, Lee, Grest, Grosberg, and
  Kremer}}]{Halverson2010a}
\bibinfo{author}{\bibfnamefont{J.~D.} \bibnamefont{Halverson}},
  \bibinfo{author}{\bibfnamefont{W.~B.} \bibnamefont{Lee}},
  \bibinfo{author}{\bibfnamefont{G.~S.} \bibnamefont{Grest}},
  \bibinfo{author}{\bibfnamefont{A.~Y.} \bibnamefont{Grosberg}},
  \bibnamefont{and} \bibinfo{author}{\bibfnamefont{K.}~\bibnamefont{Kremer}}
  (\bibinfo{year}{2010}{\natexlab{b}}).

\bibitem[{\citenamefont{Doi and Edwards}(1986)}]{Doi1986}
\bibinfo{author}{\bibfnamefont{M.}~\bibnamefont{Doi}} \bibnamefont{and}
  \bibinfo{author}{\bibfnamefont{S.}~\bibnamefont{Edwards}},
  \emph{\bibinfo{title}{{The Theory of Polymer Dynamics}}}
  (\bibinfo{publisher}{Oxford University Press}, \bibinfo{address}{Oxford},
  \bibinfo{year}{1986}).

\bibitem[{\citenamefont{Kimura and Cook}(2007)}]{Kimura2007}
\bibinfo{author}{\bibfnamefont{H.}~\bibnamefont{Kimura}} \bibnamefont{and}
  \bibinfo{author}{\bibfnamefont{P.}~\bibnamefont{Cook}}, in
  \emph{\bibinfo{booktitle}{Nuclear dynamics: molecular biology and
  visualization of the nucleus}} (\bibinfo{publisher}{Springer Japan},
  \bibinfo{year}{2007}).

\bibitem[{\citenamefont{Meaburn and Misteli}(2007)}]{Meaburn2007}
\bibinfo{author}{\bibfnamefont{K.~J.} \bibnamefont{Meaburn}} \bibnamefont{and}
  \bibinfo{author}{\bibfnamefont{T.}~\bibnamefont{Misteli}},
  \bibinfo{journal}{Nature} \textbf{\bibinfo{volume}{445}},
  \bibinfo{pages}{379} (\bibinfo{year}{2007}).

\bibitem[{\citenamefont{Dorier and Stasiak}(2009)}]{Dorier2009}
\bibinfo{author}{\bibfnamefont{J.}~\bibnamefont{Dorier}} \bibnamefont{and}
  \bibinfo{author}{\bibfnamefont{A.}~\bibnamefont{Stasiak}},
  \bibinfo{journal}{Nucleic Acids Res.} \textbf{\bibinfo{volume}{37}},
  \bibinfo{pages}{6316} (\bibinfo{year}{2009}).

\bibitem[{\citenamefont{Lieberman-Aiden
  et~al.}(2009)\citenamefont{Lieberman-Aiden, van Berkum, Williams, Imakaev,
  Ragoczy, Telling, Amit, Lajoie, Sabo, Dorschner
  et~al.}}]{Lieberman-Aiden2009}
\bibinfo{author}{\bibfnamefont{E.}~\bibnamefont{Lieberman-Aiden}},
  \bibinfo{author}{\bibfnamefont{N.~L.} \bibnamefont{van Berkum}},
  \bibinfo{author}{\bibfnamefont{L.}~\bibnamefont{Williams}},
  \bibinfo{author}{\bibfnamefont{M.}~\bibnamefont{Imakaev}},
  \bibinfo{author}{\bibfnamefont{T.}~\bibnamefont{Ragoczy}},
  \bibinfo{author}{\bibfnamefont{A.}~\bibnamefont{Telling}},
  \bibinfo{author}{\bibfnamefont{I.}~\bibnamefont{Amit}},
  \bibinfo{author}{\bibfnamefont{B.~R.} \bibnamefont{Lajoie}},
  \bibinfo{author}{\bibfnamefont{P.~J.} \bibnamefont{Sabo}},
  \bibinfo{author}{\bibfnamefont{M.~O.} \bibnamefont{Dorschner}},
  \bibnamefont{et~al.}, \bibinfo{journal}{Science}
  \textbf{\bibinfo{volume}{326}}, \bibinfo{pages}{289} (\bibinfo{year}{2009}).

\bibitem[{\citenamefont{Alberts et~al.}(2008)\citenamefont{Alberts, Johnson,
  Walter, Lewis, Raff, and Roberts}}]{Alberts2010}
\bibinfo{author}{\bibfnamefont{B.}~\bibnamefont{Alberts}},
  \bibinfo{author}{\bibfnamefont{A.}~\bibnamefont{Johnson}},
  \bibinfo{author}{\bibfnamefont{P.}~\bibnamefont{Walter}},
  \bibinfo{author}{\bibfnamefont{J.}~\bibnamefont{Lewis}},
  \bibinfo{author}{\bibfnamefont{M.}~\bibnamefont{Raff}}, \bibnamefont{and}
  \bibinfo{author}{\bibfnamefont{K.}~\bibnamefont{Roberts}},
  \emph{\bibinfo{title}{{Molecular Biology of the Cell}}}
  (\bibinfo{publisher}{Taylor \& Francis}, \bibinfo{address}{London},
  \bibinfo{year}{2008}), \bibinfo{edition}{5th} ed.

\bibitem[{\citenamefont{Witz et~al.}(2008)\citenamefont{Witz, Rechendorff,
  Adamcik, and Dietler}}]{Witz2008}
\bibinfo{author}{\bibfnamefont{G.}~\bibnamefont{Witz}},
  \bibinfo{author}{\bibfnamefont{K.}~\bibnamefont{Rechendorff}},
  \bibinfo{author}{\bibfnamefont{J.}~\bibnamefont{Adamcik}}, \bibnamefont{and}
  \bibinfo{author}{\bibfnamefont{G.}~\bibnamefont{Dietler}},
  \bibinfo{journal}{Phys. Rev. Lett.} \textbf{\bibinfo{volume}{101}},
  \bibinfo{pages}{3} (\bibinfo{year}{2008}).

\bibitem[{\citenamefont{Sanchez et~al.}(2010)\citenamefont{Sanchez, Kulic, and
  Dogic}}]{Sanchez2010}
\bibinfo{author}{\bibfnamefont{T.}~\bibnamefont{Sanchez}},
  \bibinfo{author}{\bibfnamefont{I.~M.} \bibnamefont{Kulic}}, \bibnamefont{and}
  \bibinfo{author}{\bibfnamefont{Z.}~\bibnamefont{Dogic}},
  \bibinfo{journal}{Phys. Rev. Lett.} \textbf{\bibinfo{volume}{104}},
  \bibinfo{pages}{65} (\bibinfo{year}{2010}).

\bibitem[{\citenamefont{Liu and Chakraborty}(2008)}]{Liu2008}
\bibinfo{author}{\bibfnamefont{Y.}~\bibnamefont{Liu}} \bibnamefont{and}
  \bibinfo{author}{\bibfnamefont{B.}~\bibnamefont{Chakraborty}},
  \bibinfo{journal}{Phys. Biol.} \textbf{\bibinfo{volume}{5}},
  \bibinfo{pages}{026004} (\bibinfo{year}{2008}).

\bibitem[{\citenamefont{Fritsche and Heermann}(2011)}]{Fritsche2011}
\bibinfo{author}{\bibfnamefont{M.}~\bibnamefont{Fritsche}} \bibnamefont{and}
  \bibinfo{author}{\bibfnamefont{D.~W.} \bibnamefont{Heermann}}
  (\bibinfo{year}{2011}).

\bibitem[{\citenamefont{Grosberg et~al.}(1988)\citenamefont{Grosberg, Nechaev,
  and Shakhnovich}}]{Grosberg1988}
\bibinfo{author}{\bibfnamefont{A.}~\bibnamefont{Grosberg}},
  \bibinfo{author}{\bibfnamefont{S.}~\bibnamefont{Nechaev}}, \bibnamefont{and}
  \bibinfo{author}{\bibfnamefont{E.}~\bibnamefont{Shakhnovich}},
  \bibinfo{journal}{J. Phys.} \textbf{\bibinfo{volume}{49}},
  \bibinfo{pages}{2095} (\bibinfo{year}{1988}).

\bibitem[{\citenamefont{Ostermeir et~al.}(2010)\citenamefont{Ostermeir, Alim,
  and Frey}}]{Ostermeir2010}
\bibinfo{author}{\bibfnamefont{K.}~\bibnamefont{Ostermeir}},
  \bibinfo{author}{\bibfnamefont{K.}~\bibnamefont{Alim}}, \bibnamefont{and}
  \bibinfo{author}{\bibfnamefont{E.}~\bibnamefont{Frey}},
  \bibinfo{journal}{Phys. Rev. E.} \textbf{\bibinfo{volume}{81}},
  \bibinfo{pages}{1} (\bibinfo{year}{2010}).

\bibitem[{\citenamefont{de~Gennes}(1979)}]{DeGennes1979}
\bibinfo{author}{\bibfnamefont{P.}~\bibnamefont{de~Gennes}},
  \emph{\bibinfo{title}{{Scaling Concepts in Polymer Physics}}}
  (\bibinfo{publisher}{Cornell University Press}, \bibinfo{address}{Ithaca},
  \bibinfo{year}{1979}).

\bibitem[{\citenamefont{Kremer and Grest}(1990)}]{Kremer1990}
\bibinfo{author}{\bibfnamefont{K.}~\bibnamefont{Kremer}} \bibnamefont{and}
  \bibinfo{author}{\bibfnamefont{G.~S.} \bibnamefont{Grest}},
  \bibinfo{journal}{J. Chem. Phys.} \textbf{\bibinfo{volume}{92}},
  \bibinfo{pages}{5057} (\bibinfo{year}{1990}).

\bibitem[{\citenamefont{Anderson et~al.}(2008)\citenamefont{Anderson, Lorenz,
  and Travesset}}]{Anderson2008}
\bibinfo{author}{\bibfnamefont{J.}~\bibnamefont{Anderson}},
  \bibinfo{author}{\bibfnamefont{C.}~\bibnamefont{Lorenz}}, \bibnamefont{and}
  \bibinfo{author}{\bibfnamefont{A.}~\bibnamefont{Travesset}},
  \bibinfo{journal}{J. Comput. Phys.} \textbf{\bibinfo{volume}{227}},
  \bibinfo{pages}{5342} (\bibinfo{year}{2008}).

\bibitem[{\citenamefont{Soddemann et~al.}(2003)\citenamefont{Soddemann,
  D\"{u}nweg, and Kremer}}]{Soddemann2003}
\bibinfo{author}{\bibfnamefont{T.}~\bibnamefont{Soddemann}},
  \bibinfo{author}{\bibfnamefont{B.}~\bibnamefont{D\"{u}nweg}},
  \bibnamefont{and} \bibinfo{author}{\bibfnamefont{K.}~\bibnamefont{Kremer}},
  \bibinfo{journal}{Phys. Rev. E.} \textbf{\bibinfo{volume}{68}},
  \bibinfo{pages}{1} (\bibinfo{year}{2003}).

\bibitem[{\citenamefont{Egorov}(2008)}]{Egorov2008}
\bibinfo{author}{\bibfnamefont{S.~A.} \bibnamefont{Egorov}},
  \bibinfo{journal}{J. Chem. Phys.} \textbf{\bibinfo{volume}{129}},
  \bibinfo{pages}{064901} (\bibinfo{year}{2008}).

\end{thebibliography}
\end{document}